\documentclass[aps,prl,reprint,amsmath,amssymb,superscriptaddress]{revtex4-2}
\makeatletter
\let\frontmatter@footnote@produce\frontmatter@footnote@produce@footnote
\@booleanfalse\footinbib@sw
\makeatother
\usepackage[T1]{fontenc}
\usepackage{newtxtext,newtxmath}
\usepackage{graphicx}
\begin{document}

\title{Electron Correlations and Electron–Phonon Coupling Lock Lattice-Spin-Charge Ordered Structures in La$_3$Ni$_2$O$_7$}

\author{Ye Liu}
\affiliation{Key Laboratory of Materials Physics, Institute of Solid State Physics, HFIPS, Chinese Academy of Science, Hefei 230031, China}
\affiliation{Science Island Branch of the Graduate School, University of Science and Technology of China, Hefei 230026, China}
\author{Shu-Hong Tang}
\affiliation{Key Laboratory of Materials Physics, Institute of Solid State Physics, HFIPS, Chinese Academy of Science, Hefei 230031, China}
\affiliation{Science Island Branch of the Graduate School, University of Science and Technology of China, Hefei 230026, China}
\author{Ya-Min Quan}
\affiliation{Key Laboratory of Materials Physics, Institute of Solid State Physics, HFIPS, Chinese Academy of Science, Hefei 230031, China}
\author{Xiao-Di Liu}
\email[Contact author: ]{xiaodi@issp.ac.cn}
\affiliation{Key Laboratory of Materials Physics, Institute of Solid State Physics, HFIPS, Chinese Academy of Science, Hefei 230031, China}
\affiliation{Science Island Branch of the Graduate School, University of Science and Technology of China, Hefei 230026, China}
\author{Liang-Jian Zou}
\email[Contact author: ]{zou@theory.issp.ac.cn}
\affiliation{Key Laboratory of Materials Physics, Institute of Solid State Physics, HFIPS, Chinese Academy of Science, Hefei 230031, China}
\affiliation{Science Island Branch of the Graduate School, University of Science and Technology of China, Hefei 230026, China}

\begin{abstract}
Utilizing the finite-temperature cluster rotationally invariant slave-boson (RISB) method, we study the ground-state properties of ambient-pressure La$_3$Ni$_2$O$_7$ with intersite Coulomb repulsion and electron–phonon (e-ph) coupling, in addition to two-orbital Hubbard correlations. We show that the electron correlations give rise to a spin stripe with charge disproportionation, and that the e-ph coupling drives the Amam structure to distort into two nearly degenerate phases: lattice distortions of the same or opposite sign in adjacent bilayers produce polar Am2m and antipolar Pmmn, respectively. The spin stripe is site-centered, with large magnetic moments and excess electrons on half of the Ni rows and nearly vanishing moments on the other half. The low-temperature full energy gap lies in $x=d_{x^2-y^2}$-dominated bands, whereas the flat $z=d_{3z^2-r^2}$ bonding bands lie below the Fermi energy. The two $z$ electrons of each interlayer dimer form a superposition of bonding-pair and antibonding-pair configurations, which requires exact diagonalization within the dimer. The energy gap grows linearly with the charge modulation beyond a threshold and, upon warming, closes continuously, leaving a metallic stripe state with both spin and charge modulation. The charge modulation linearly drives a breathing distortion of the NiO$_6$ octahedra, and the distortion, together with the intersite Coulomb repulsion, amplifies the charge modulation. Interestingly, we find that the charge order parameter is linearly proportional to the square of the spin polarization of the $x$ orbital, without requiring a Fermi-surface instability or a fully gapped state. Strong magnetic moment contrast and charge disproportionation, the partially polarized interlayer $z$-orbital dimer, and the different orbital origins of the moments and low-energy spectrum, can thus be connected in one physical scenario.
\end{abstract}

\maketitle

The discovery of unconventional superconductivity in bilayer Ruddlesden–Popper (RP) La$_3$Ni$_2$O$_7$ with critical temperature higher than the boiling point of liquid nitrogen provides a new strongly correlated platform, in addition to cuprate compounds \cite{Sun2023Nature,Hou2023CPL,ZhangYN2024NatPhys,WangG2024PRX,WangN2024Nature,LiF2026Nature}. With the deepening understanding of the microscopic electronic properties of bilayer nickelate La$_3$Ni$_2$O$_7$, which displays superconductivity under hydrostatic pressure, one realizes that the reported normal-state properties of strongly correlated La$_3$Ni$_2$O$_7$ are far from well consistent in crystalline structure, ground-state magnetic configuration, interlayer coupling, as well as underlying orders of spin and charge density waves, etc. \cite{WangM2024CPL}. Exactly and consistently understanding these properties is crucial for uncovering the superconductive origin of pressurized La$_3$Ni$_2$O$_7$ and related nickelates.

In the crystalline structure aspect, earlier experimental observations \cite{Sun2023Nature,ref45,ref13,ChenXL2024JACS} and theoretical calculations \cite{ref18,Geisler2024npjQM,Rhodes2024PRM} suggested that ambient- and low-pressure La$_3$Ni$_2$O$_7$ adopts the Amam phase, as shown in Fig.~1(b); however, recent single-crystal diffraction experiments \cite{ref05,Sasaki2026arXiv,ref15,ref07} and the density-functional-theory plus $U$ (DFT+$U$) calculation \cite{ref21} proposed that the polar Am2m phase shown in Fig.~1(a), rather than the Amam phase, is the ground-state structure of bilayer La$_3$Ni$_2$O$_7$, and a hybrid-functional calculation obtained a polar structure of the same space group \cite{Foyevtsova2025PRB}. On the other hand, Zhang et al. \cite{ZhangB2025PRB} found that the most favorable crystalline structure in ambient-pressure La$_3$Ni$_2$O$_7$ was the Pmnm phase [Pmmn in the setting used here, Fig.~1(c)] by utilizing the Perdew–Burke–Ernzerhof plus $U$ (PBE+$U$) method; distorted phases in the same space groups as Am2m and Pmnm, as well as Pnma and P2$_1$/m, also appeared in the calculation of Ref.~\cite{ref17}. These debated and conflicting results on the crystalline structure in ambient-pressure La$_3$Ni$_2$O$_7$, together with the reported stacking polymorphs \cite{ChenXL2024JACS,Puphal2024PRL} and a monoclinic assignment \cite{LiF2026Nature}, call for not only further refined experimental identification but also more accurate theoretical exploration.

\begin{figure*}[p]
\centering
\includegraphics[width=0.946\textwidth]{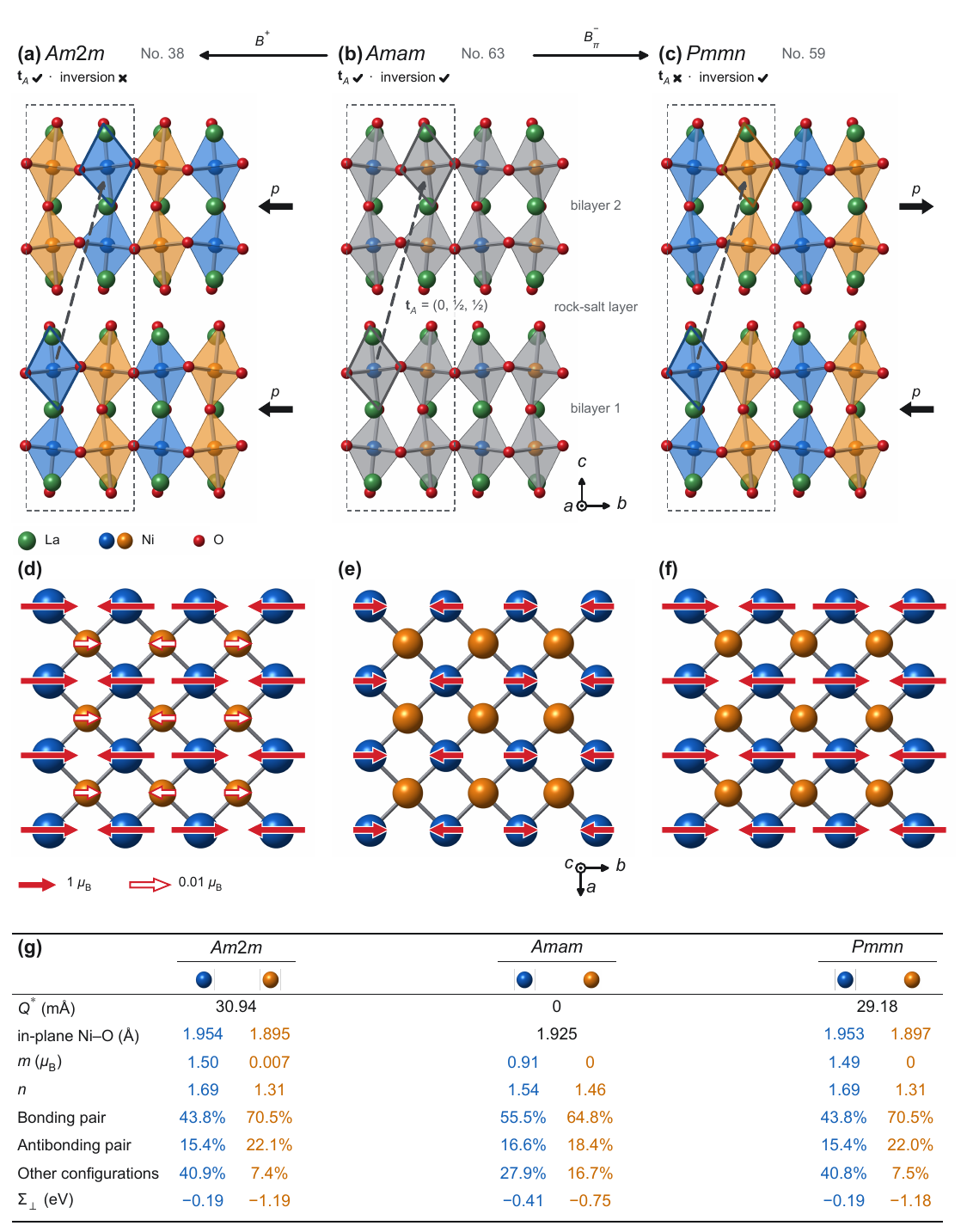}
\caption{Site-centered spin stripes at $T=0$ in polar Am2m (a), undistorted Amam (b), and antipolar Pmmn (c). Blue (orange) marks the same electron-rich, large-moment (nearly nonmagnetic) Ni in the lattice (a)–(c), in one Ni layer (d)–(f), and in the table (g); its octahedron is expanded (contracted) in (a) and (c). Dashed arrows: A-centering translation $\mathbf t_A$ relating the outlined octahedra; $p$: bilayer dipole direction. (g) lists breathing amplitude $Q^*$, bond lengths, moment $m$, occupation $n$, $d_{z^2}$ pair-configuration weights, and interlayer self-energy $\Sigma_\perp$.}
\end{figure*}

Magnetic order in the ambient-pressure bilayer nickelate La$_3$Ni$_2$O$_7$ involves electronic degrees of freedom at different levels. Many efforts have been made to clarify the ground-state magnetic moment and magnetic configuration of Ni spins in ambient-pressure La$_3$Ni$_2$O$_7$ \cite{WangY2024PRB,ref20,Ni2025npjQM}; various magnetic moments and configurations were presented. Theoretically, on the basis of the ambient Amam structure, Wang et al. \cite{WangY2024PRB} showed the presence of an entangled spin and charge double stripe ground state within the mean-field method and DFT+$U$ technique; meanwhile, LaBollita et al. \cite{ref20} predicted the ground state with alternating high-low Ni spins with double stripe and single spin-charge stripe, and the ground state is insulating at $U=0$ eV; however, Ni et al. \cite{Ni2025npjQM} suggested stoichiometric La$_3$Ni$_2$O$_7$ has a double striped magnetic ground state. It remains unknown how the ground-state magnetic configuration changes when the stable crystalline structure varies from the Amam to the Am2m or Pmmn phase.

Experimentally, single-crystal neutron scattering provides indications of spin stripe order in the NiO$_2$ plane and strong antiferromagnetic interlayer coupling \cite{ref01}, consistent with the strong interlayer exchange found by resonant inelastic x-ray scattering (RIXS) \cite{ChenXY2024NC} and inelastic neutron scattering \cite{Xie2024SciBull}. Neutron powder diffraction \cite{ref02} and nuclear quadrupole resonance (NQR) reanalyzed using the Am2m structure \cite{ref03} point to alternating Ni environments with large and small moments, as do the muon-spin rotation/relaxation ($\mu$SR) \cite{ref46} and Raman scattering \cite{Gim2026PRX} experiments. On the other hand, with the neutron scattering experiment, Xie et al. \cite{Xie2024SciBull} observed weak and flat spin fluctuations at about 45 meV and absence of long-range magnetic order down to 10 K in the polycrystal. However, Chen et al. \cite{ref01} showed single stripe order with the wave vector $\mathbf q = (0, 0.5, 0.5)$, finite spin gap, and obvious bilayer modulation with interlayer exchange coupling $SJ_\perp \approx 40$ meV $\gg SJ_\parallel$, with $S$ the spin magnitude. With the RIXS technique, Chen et al. \cite{ChenXY2024NC} showed that Ni $3d_{x^2-y^2}$, $3d_{3z^2-r^2}$ and O $2p$ orbitals dominate the low-energy physics, and the interlayer exchange $J_\perp \gg J_\parallel$; Gupta et al. \cite{ref04} found no evidence for charge or bond disproportionation in the resonant soft x-ray scattering of the epitaxial film over 20 K to 200 K.

Through the $\mu$SR technique, Chen et al. \cite{ref06} found bulk magnetic order at 154 K, and no anomaly in the resistivity and magnetic susceptibility in 100–200 K; Khasanov et al. \cite{ref46} suggested an alternating moment-stripe and momentless stripe magnetic ground state. With increasing pressure, the magnetic transition temperature lifts, while another density-wave transition temperature decreases, implying the latter to be a charge-density-wave (CDW) transition. Plokhikh et al. \cite{ref02} found that in the Cmcm structure (Amam in the setting used here) the in-plane Ni spins display an alternating distribution with small moment ($m=0.05$–$0.075\,\mu_{\rm B}$) and large moment ($m=0.66\,\mu_{\rm B}$). Yashima et al.'s $^{139}$La NQR spectrum analysis \cite{ref03} showed that in the Am2m structure the spin-density-wave (SDW) magnetic configuration displays two inequivalent Ni spins of large and micro magnetic moments. Gim et al. \cite{Gim2026PRX} suggested spin disproportionation in the ambient-pressure Raman scattering experiment. Misawa et al.'s work \cite{ref05} also supported two kinds of Ni sites.

Low-temperature transport and optical anomalies can be separated from the onset of magnetic order \cite{ref13,ref15,ref46,ref14}. Liu et al. \cite{ref13} showed that the resistivity of the single crystal displays two anomalies at $T=110$ K and 153 K, respectively. The former only appears in resistivity, while the latter also appears in specific heat and magnetic susceptibility. Wu et al. \cite{ref14} identified through the ultrafast pump–probe that there exists an electronically nematic SDW phase below 140 K, and a nonmagnetic CDW phase below 115 K; the latter is completely suppressed by very weak pump, while the former persists to strong pump.

Comparing the tight-binding parameters of nickelates with those of cuprates, one finds an outstanding character of bilayer nickelate superconductors differing from cuprate superconductors is that bilayer Ni $3d_{3z^2-r^2}$ orbitals strongly hybridize and interact, forming a dimer unit. Khaliullin and Chaloupka proposed magnetic instability on a singlet background \cite{ref33}, and the competition of the Hund's coupling and superexchange interaction contributes high- and low-spins, and transfers the spin polarization of the bilayer Ni $3d_{x^2-y^2}$ orbitals to the singlets of interlayer Ni $3d_{3z^2-r^2}$ orbitals, as suggested in Refs.~\cite{OhZhang2023PRB,Oh2026PRB}. Recently Lechermann et al. applied two-Ni cluster rotationally invariant slave bosons (RISB) to the high-pressure paramagnetic state of this material \cite{ref34}. Very recently LaBollita et al. utilizing the cluster dynamical mean-field theory \cite{LaBollita2026arXiv} and Liao et al. using the slave spin method \cite{Liao2026PRB} found the interlayer $3d_{3z^2-r^2}$ orbitals form singlets. Experimentally, Lee et al. \cite{ref08} showed that in the nuclear magnetic resonance (NMR) measurement the interlayer $3d_{3z^2-r^2}$ orbitals form a spin singlet, coexisting with the SDW in $3d_{x^2-y^2}$ orbitals. Nevertheless, whether the two Ni ions form an $S=0$ singlet or a partially polarized dimer deserves further investigation, especially considering the possible double-striped spin structure.

Experimentally, Liu et al. \cite{ref15} identified two density waves which behave with opposite pressure dependence: the high-$T$ one is associated with SDW. The essence of the low-$T$ one is unknown. In the NQR measurement Luo et al. found that the charge modulation forms together with the spin order \cite{LuoJ2025CPL}. Khasanov et al. \cite{ref46} found that increasing pressure lifts the magnetic order transition temperature, while another lower density-wave order transition temperature declines, which might be corresponding to the CDW order.

Theoretically, linear–quadratic coupling between spin order and charge order at twice its wave vector had precedents in the Landau theory \cite{ref36,ref37,Norman2025PRB}, and related discussion has appeared for this material \cite{ZhangB2025PRB,Ni2025npjQM,ref32}. In their zero-temperature mean-field study, Deng et al. obtained intertwined double spin-charge stripes with charge order driven by intersite repulsion; their model contains neither lattice degrees of freedom nor many-body correlations within a dimer \cite{ref28}. Cluster exact diagonalization \cite{JiangG2025SCPMA} and unrestricted Hartree–Fock \cite{ref29} also give intertwined charge and spin orders. Nearest-neighbor repulsion is a nonlocal Coulomb interaction that acts directly on the charge channel: correlated-screening calculations for high-pressure structures and models of nonlocal Coulomb repulsion have examined tendencies toward charge order \cite{ref26,ref27,Du2025APS}. Zhan et al. \cite{ref27} using the functional renormalization group technique showed that the interlayer intra-orbital repulsion promotes the charge order breaking the mirror symmetry of two Ni layers, while the intralayer nearest-neighbor repulsion favors the in-plane CDW with $\mathbf q=(\pi, \pi)$.

Foyevtsova et al.'s hybrid functional study stressed the nonlocal correlation effects arising in this system from the molecular orbital splitting of the Ni $3d_{3z^2-r^2}$ states inside Ni–Ni dimers \cite{Foyevtsova2025PRB}. Utilizing the unrestricted Hartree–Fock approximation, Braz et al. found that La$_3$Ni$_2$O$_7$ can produce moment differentiation along $(0, \pi)$ and accompanying charge modulation \cite{ref29}; the charge order redistributes the magnetic moments, generating low-spin sites, and the charge-order parameter is one order smaller in magnitude than the spin order parameter. In their ambient-pressure PBE+$U$ calculation, Ni et al. \cite{Ni2025npjQM} showed that in the low-symmetric magnetically ordered phase, the spin–lattice coupling may induce CDW order; Zhang et al. \cite{ZhangB2025PRB} found that in the double stripe phase, the phonon spectrum in the ambient-pressure Amam phase displays negative/imaginary frequency, and La$_3$Ni$_2$O$_7$ distorts to the Pmnm phase, accompanied by the CDW order. These dispersive results call for further and deeper research on the microscopic origin of the charge order or CDW order.

From various electronic structure calculations \cite{Liao2026PRB,JiangG2025SCPMA,ref22} one now believes that similar to cuprate superconductors, nickelate superconductors are strongly correlated systems, in which the on-site Coulomb correlation $U$ is believed to be about 5–8 eV. Angle-resolved photoemission spectroscopy (ARPES) reveals a strongly correlated $z$ flat band \cite{ref09}, suppressed spectral weight in the $x$-dominated states near the Fermi energy \cite{ref10}, and band folding with local gap opening \cite{ref11}. In ARPES, Yang et al. \cite{ref09} and Li et al. \cite{ref10} suggested that near the Brillouin-zone corner the flat band dominated by $3d_{3z^2-r^2}$ orbitals displays the strongest electron correlations. Z. Liu et al. \cite{ref12} found that in the infrared spectroscopy study the transition at 115 K removes the Drude component displaying non-Fermi liquid behavior, demonstrating strong correlation character. On the other hand, different from cuprate superconductors, two $e_g$ orbitals are involved in the superconducting pairing in nickelate compounds \cite{ref21,ref20,JiangG2025SCPMA,ref22,ref19}.

To date the importance of the electron–phonon (e-ph) coupling in ambient-pressure La$_3$Ni$_2$O$_7$ has not received enough attention. The robust phase transition induced by hydrostatic pressure \cite{Sun2023Nature,Sasaki2026arXiv,WangL2024JACS} implies that the electron–phonon coupling plays an essential role in the normal-state properties of ambient-pressure La$_3$Ni$_2$O$_7$, and hence should be taken into account. It is known that breathing modes connect charge inequivalence to the lattice \cite{Mochizuki2018PRM}. Oxygen-isotope substitution mainly shifts the lower-temperature resistive anomaly, leaving the magnetic transition temperature unchanged within experimental uncertainty \cite{ref16}; Raman scattering reported $B_{1g}$ phonons softening from 280 K down to 4.5 K \cite{Gim2026PRX}, whereas another Raman study saw no phonon anomaly across the transition \cite{He2026NC}. Theoretically, Yi et al. obtained phonon instabilities and breathing distortions on a magnetically ordered background in the DFT+$U$ framework \cite{ref21,ZhangB2025PRB,ref17}, whereas the phonon calculations including $U$ by Zhang et al. gave dynamical stability of the undistorted Amam parent structure \cite{ref18}.

These inconsistent observations suggest that ordered magnetic moments, charge disproportionation, and structural inequivalence should be considered on the same footing, as in the site-selective Mott phase of the perovskite nickelates \cite{Park2012PRL}. These previous studies motivate us to examine the cooperation of magnetism, charge and the lattice; the specific causal relationships must be established by energy comparisons and constrained controls within the same model. Here we retain on-site and intersite Coulomb repulsion, local Hund exchange, dimer correlations, and structural degrees of freedom together, using cluster RISB with electron–phonon coupling. Because the anomalies discussed above occur upon warming, zero-temperature total energies alone are insufficient; we adopt a finite-temperature scheme \cite{ref35} with the Weiss temperature scale anchored to the experimental magnetic transition temperature. We find that the electron correlations lead to the spin stripe order and the charge disproportionation, and that the e-ph coupling leads to two nearly degenerate structural minima with different distortions, stabilizing the ground state as polar Am2m or antipolar Pmmn phases, respectively; the two $z=d_{3z^2-r^2}$ electrons of each pair of interlayer Ni ions form a dimer with partial spin polarization; and the charge order parameter is linearly proportional to the square of the spin polarization of the $x=d_{x^2-y^2}$ orbital, etc. These results systematically address various properties of ambient La$_3$Ni$_2$O$_7$, establishing a milestone for further understanding the microscopic origin of superconducting La$_3$Ni$_2$O$_7$.

\textit{Model---}The model Hamiltonian describing the low-energy processes of ambient La$_3$Ni$_2$O$_7$ consists of the electron part, their coupling to the phonon, and the phonon part,
\begin{equation}
H=H_{\rm el}+H_{\rm e\text{-}ph}+H_{\rm lat}. \tag{1}
\end{equation}

For the electrons we use the bilayer two-orbital model \cite{ref39} of the Ni $e_g$ orbitals $x=d_{x^2-y^2}$ and $z=d_{3z^2-r^2}$,
\begin{equation}
H_{\rm el}=\sum_{ij,ab,\sigma}t^{ab}_{ij}d^\dagger_{ia\sigma}d_{jb\sigma}+\sum_{ia}\epsilon_a n_{ia}+\sum_i H_U(i)+H_V, \tag{2}
\end{equation}
with hoppings and level energies fitted to the nonmagnetic bands of the undistorted Amam parent structure, at 1.5 electrons per Ni. $H_U$ is the full Kanamori interaction \cite{ref40,ref41}, including spin-flip and pair-hopping terms, with the representative values $U=5$ eV and $J_H=0.5$ eV. The intersite repulsion acts between in-plane and vertical nearest neighbors, $\ell=\parallel,\perp$,
\begin{equation}
H_V=\sum_{\ell}\sum_{\langle ij\rangle_\ell}\sum_{ab}V_\ell^{ab}\,\delta n_{ia}\,\delta n_{jb}, \tag{3}
\end{equation}
where $\delta n_{ia}=n_{ia}-\bar n_a$ is the occupation measured from a fixed uniform background, $\bar n_x=0.54$ and $\bar n_z=0.96$, the site-averaged orbital occupations of a stripe solution on the undistorted lattice. The six elements $V_\ell^{ab}$ are Coulomb integrals between the Wannier charge densities of orbitals $a$ and $b$ on neighboring Ni, evaluated with a Thomas–Fermi-screened kernel. The screening length sets both their ratios and their overall strength $V_{\rm rep}=(4\bar V_\parallel+\bar V_\perp)/5$, where $\bar V_\ell$ is the orbital average of $V_\ell^{ab}$; we take $V_{\rm rep}=0.15$ eV.

The lattice part enters through six frozen displacement patterns with amplitudes $Q_\nu$, each measured by the largest atomic displacement of its pattern. Two of them are the in-plane breathing of the NiO$_6$ octahedra shown in Figs.~1(a)–1(c): the octahedra on one checkerboard sublattice of each layer expand and those on the other contract, alike in the two layers of a bilayer. The A-centering translation $\mathbf t_A=(0,\tfrac12,\tfrac12)$ of the parent lattice maps each bilayer onto the next, and the breathing of adjacent bilayers, compared at positions related by $\mathbf t_A$, has either the same sign or opposite signs. The in-phase breathing mode $B^+$ is thus even under $\mathbf t_A$ and the antiphase breathing mode $B_\pi^-$ odd; we denote their amplitudes by $Q_{B^+}$ and $Q_{B_\pi^-}$. The other four modes are a breathing pattern of opposite sign in the two layers of a bilayer, $D^-$, and three patterns in which the inner-apical ligands move along the $c$ axis, $A^-$, $C_0^+$, and $C_\pi^-$. Each pattern shifts the on-site energies linearly; this is the electron–phonon coupling,
\begin{equation}
H_{\rm e\text{-}ph}=-\sum_{\nu}Q_\nu\sum_{ia}\Lambda_{a\nu}\,\eta_i^{(\nu)}\,n_{ia}, \tag{4}
\end{equation}
where $\eta_i^{(\nu)}=\pm1$ is the site pattern of mode $\nu$. The displacements thus act on the electrons only through the on-site energies; the hopping amplitudes $t^{ab}_{ij}$ of Eq.~(2) keep the values of the undistorted structure. The lattice carries the harmonic elastic energy
\begin{equation}
H_{\rm lat}=\frac{N_{\rm Ni}}{2}\sum_{\nu\nu'}Q_\nu K_{\nu\nu'}Q_{\nu'}. \tag{5}
\end{equation}

The zero of energy is taken at the undistorted Amam parent structure, $Q=0$, where $H_{\rm e\text{-}ph}$ and $H_{\rm lat}$ both vanish; the energies of the distorted structures are measured from the solution on this undistorted lattice. The deformation potentials $\Lambda$ and stiffnesses $K$ are obtained from frozen-displacement density-functional (PBE) \cite{Perdew1996PRL,Blaha2020JCP} calculations for the nonmagnetic parent phase. The breathing modes couple almost entirely to the $x$ orbital: $\Lambda_{x\nu}=4.45$ (4.42) eV/Å and $\Lambda_{z\nu}=-0.25$ ($-0.20$) eV/Å for $\nu=B^+$ ($B_\pi^-$), with $K_{\nu\nu}=33.85$ (35.06) eV/Å$^2$ per Ni.

\textit{Method---}We take the vertical Ni–Ni dimer as the local unit and diagonalize its complete many-body Hamiltonian exactly in the 256-dimensional Fock space, keeping all particle-number sectors. The vertical hopping $T_\perp$ (with amplitude $t_\perp$ between the two $z$ orbitals), the Kanamori interaction, and the vertical repulsion are thereby treated exactly; the in-plane repulsion enters as a Hartree potential. With the electron–phonon coupling of Eq.~(4), the local Hamiltonian of a dimer is
\begin{equation}
\begin{aligned}[b]
H_{\rm loc}&=T_\perp+H_V^{\perp}+\sum_{i}\Bigl[H_U(i)+\sum_a(\epsilon_a+v_{ia})\,n_{ia}\Bigr],\\
v_{ia}&=\sum_{j,b}V_\parallel^{ab}\langle\delta n_{jb}\rangle-\sum_\nu\Lambda_{a\nu}\eta_i^{(\nu)}Q_\nu,
\end{aligned} \tag{6}
\end{equation}
where $i$ runs over the two Ni of the dimer, $j$ over the in-plane nearest neighbors of $i$, and $H_V^\perp$ is the vertical term of Eq.~(3) on this dimer. Motion between dimers is treated with cluster RISB \cite{ref34,ref31,ref43}, whose quasiparticle Hamiltonian is
\begin{equation}
h_{\rm qp}(\mathbf k)=R^\dagger[h_{\rm hop}(\mathbf k)-T_\perp]R+\mathcal L. \tag{7}
\end{equation}

Here the renormalization matrix $R$ and the Lagrange multipliers $\mathcal L$ follow from the saddle-point equations, and local observables are evaluated with the dimer density matrix of the same saddle point. The interlayer self-energy of the $z$ orbital at zero frequency is $\Sigma_\perp=\tfrac12\sum_\sigma[R^{-\dagger}(\mathcal L-\mu)R^{-1}-T_\perp]_{tz\sigma,bz\sigma}$, where $t$ and $b$ label the two layers.

The electron part and the lattice part are solved together. At each $Q$ the electronic problem is converged with the level shifts of Eq.~(6), giving the electronic free energy per Ni, $F_{\rm el}(Q)$. The free energy measured from the undistorted solution,
\begin{equation}
\mathcal F(Q)=F_{\rm el}(Q)-F_{\rm el}(0)+\frac12\sum_{\nu\nu'}Q_\nu K_{\nu\nu'}Q_{\nu'}, \tag{8}
\end{equation}
is minimized over all six coordinates, at every temperature and every $V_{\rm rep}$. Its stationary points obey the force balance
\begin{equation}
\begin{aligned}[b]
\sum_{\nu'}K_{\nu\nu'}Q_{\nu'}&=\sum_a\Lambda_{a\nu}\rho_{a\nu},\\
\rho_{a\nu}&=\frac1{N_{\rm Ni}}\sum_i\eta_i^{(\nu)}\langle n_{ia}\rangle.
\end{aligned} \tag{9}
\end{equation}

The loop is closed: $Q$ shifts the levels through Eq.~(6), and the charge distribution of the resulting solution fixes $Q$ through Eq.~(9). For the in-phase and antiphase breathing modes, $\nu=B^+$ and $B_\pi^-$, Eq.~(9) gives
\begin{equation}
Q_\nu=\frac{1}{2K_{\nu\nu}}\sum_a\Lambda_{a\nu}\,\Delta n_a, \tag{10}
\end{equation}
where $\Delta n_a$ is the difference in the occupation of orbital $a$ between the Ni on expanded and contracted octahedra, and the electron–phonon and elastic terms together amount to
\begin{equation}
\langle H_{\rm e\text{-}ph}\rangle+H_{\rm lat}=-\frac{N_{\rm Ni}}{8K_{\nu\nu}}\Bigl(\sum_a\Lambda_{a\nu}\,\Delta n_a\Bigr)^2, \tag{11}
\end{equation}
an attraction in the charge channel that lowers the cost of the charge contrast.

Because the saddle point carries no entropy of collective spin fluctuations, the thermal evolution of the ordered moment is supplied by the Weiss self-consistent temperature scale, anchored solely to the experimental magnetic transition temperature $T_{\rm SDW}=150$ K \cite{ref08}:
\begin{equation}
M(T)=M_0\lambda(T),\qquad\lambda=\tanh\!\left(\lambda\,T_{\rm SDW}/T\right), \tag{12}
\end{equation}
where $M$ is the moment of the large-moment rows and $M_0$ its zero-temperature value. Only $M$ is prescribed; its partition between the two orbitals, the charge distribution, and all six structural coordinates are solved self-consistently at every temperature, with the thermal occupations and entropy of the quasiparticles included as in the finite-temperature functional of Ref.~\cite{ref35}.

In our calculations the supercell holds two bilayers with 16 Ni. The magnetic moments are collinear, and the only symmetry imposed is a translation by half the magnetic period combined with spin reversal. The magnetic moments and electron occupations of the inequivalent sites and orbitals, including those of the two Ni in each dimer, are free.

\textit{Two structural minima---}Within the collinear period-four supercell, the same Hamiltonian has two nearly degenerate structural minima, shown in Fig.~2: polar Am2m and antipolar Pmmn, 9.25 and 8.48 meV/Ni below the undistorted solution and separated by a barrier of 4–5 meV/Ni. Although all six structural coordinates are relaxed, only one breathing amplitude, denoted $Q^*$, is finite at each minimum ($Q_{B^+}$ in Am2m and $Q_{B_\pi^-}$ in Pmmn), the other five remaining below $10^{-3}$ mÅ. As shown in Figs.~1(a)–1(c), breathing of equal sign in adjacent bilayers ($B^+$) keeps $\mathbf t_A$ but breaks inversion (Am2m); opposite signs ($B_\pi^-$) do the reverse (Pmmn). The charge contrast between the two kinds of Ni in Fig.~1 linearly drives the breathing distortion, expanding the electron-rich octahedra. Through the distortion, electron–phonon coupling amplifies the charge order nearly fivefold: the charge contrast is 0.08 in the stripe on the undistorted lattice and about 0.38 at either minimum, as the occupations listed in Fig.~1(g) show. The Am2m phase is lower because $B^+$ is softer and couples slightly more strongly to the $x$ charge. The pair persists down to $V_{\rm rep}=0.02$ eV.

\begin{figure}[t]
\centering
\includegraphics[width=\columnwidth]{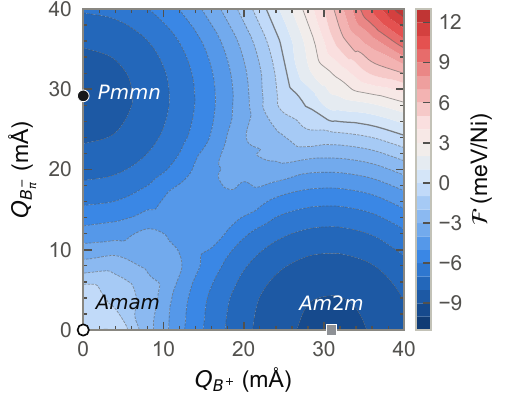}
\caption{Free energy at $T=0$ versus the in-phase and antiphase breathing amplitudes, for the lower-energy stacking. Symbols mark the two minima and the undistorted structure.}
\end{figure}

\textit{Site-centered spin stripe---}Both minima host the site-centered stripe shown in Figs.~1(d) and 1(f): rows of large-moment Ni alternate with nearly nonmagnetic rows, with moments reversed on successive large-moment rows and across each dimer, and with the large-moment rows of adjacent bilayers displaced by one row in Am2m and aligned in Pmmn, the stackings of the magnetic orders with $\mathbf q=(0,0.5,0.5)$ and $(0,0.5,0)$, respectively. In the nomenclature of Ref.~\cite{ref20}, this is the single spin-charge stripe (up/0/down/0), as distinct from the double spin stripe (up/up/down/down). It corresponds to the single-stripe order with antiferromagnetic coupling between the two layers of a bilayer found by single-crystal neutron scattering \cite{ref01}; the nearly nonmagnetic rows correspond to the nonmagnetic stripes of that analysis and of $\mu$SR \cite{ref46}, and to the small-moment sites inferred from neutron powder diffraction \cite{ref02} and NQR \cite{ref03}. The moments are listed in Fig.~1(g): the large moment, about $1.5\,\mu_{\rm B}$, is shared by the $x$ and $z$ orbitals, roughly 0.71 and 0.78 $\mu_{\rm B}$, and the other rows carry $0.007\,\mu_{\rm B}$ in Am2m and none in Pmmn. On the undistorted lattice of Fig.~1(e) the stripe is already present but weaker, with $0.91\,\mu_{\rm B}$.

\textit{Partially polarized Ni–Ni dimer---}As expected, the two $z$ electrons of a dimer, shown in Fig.~3(a), form a superposition of bonding-pair and antibonding-pair configurations, with the weights listed in Fig.~1(g). The antibonding-pair weight is 22\% on the nearly nonmagnetic rows and 15\% on the large-moment rows, so exact diagonalization within the dimer is required. Correspondingly, $\Sigma_\perp\approx-1.2$ eV on the nearly nonmagnetic rows is more than twice the bare hopping $t_\perp=-0.55$ eV. The two pair configurations carry over 92\% of the weight there, against 59\% on the large-moment rows. On the large-moment rows the pair is partially polarized: the two Ni of a dimer carry opposite $z$ moments of 0.78 $\mu_{\rm B}$, configurations with one electron in the bonding and one in the antibonding orbital take 25\% of the weight, and $\Sigma_\perp$ is reduced to $-0.19$ eV. On the nearly nonmagnetic rows the pair is unpolarized, and such configurations take 3\%. In the theory of Khaliullin and Chaloupka \cite{ref33}, local $z$ spins are bound into interlayer singlet pairs, and magnetic order arises by triplon condensation, the triplon field being the difference of the two spins of a dimer. The unpolarized pair of the nearly nonmagnetic rows corresponds to their singlet background and the partially polarized pair of the large-moment rows to their condensate; a singlet of two local spins has equal bonding-pair and antibonding-pair weights, whereas here the bonding pair carries about three times the weight of the antibonding pair.

\begin{figure}[t]
\centering
\includegraphics[width=\columnwidth]{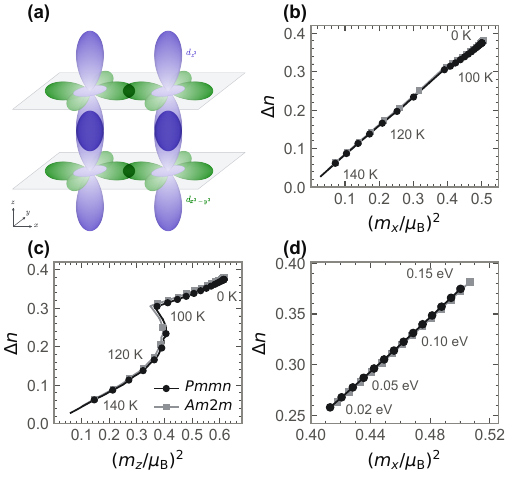}
\caption{(a) $e_g$ orbitals on two dimers: $d_{z^2}$ (purple), $d_{x^2-y^2}$ (green). $\Delta n$ versus (b) $m_x^2$ and (c) $m_z^2$ with varying temperature; (d) versus $m_x^2$ with varying $V_{\rm rep}$ at $T=0$.}
\end{figure}

\textit{Charge order versus the $x$ moment---}Charge is even and the moment odd under time reversal, so the stripe couples to the charge contrast $\Delta n$ between the two kinds of rows only through the square of the moment. On the site-centered stripe this square alternates from row to row, finite on the large-moment rows and zero on the others, which is exactly the period-two pattern of $\Delta n$ \cite{ref32}. Minimizing $F=\tfrac12A_c\Delta n^2-g_c\Delta n\,u^2$, with $u=m_x/\mu_{\rm B}$ the $x$ moment of the large-moment rows, gives $\Delta n=(g_c/A_c)u^2$. The $x$ moment enters because the charge moves between rows through the in-plane $x$ orbitals shown in Fig.~3(a), the $z$ occupation staying near one on all rows. Accordingly, Figs.~3(b) and 3(d) show that $\Delta n$ is linear in $u^2$ on warming and under varying $V_{\rm rep}$, respectively, whereas in Fig.~3(c) it is not a function of $m_z^2$.

\textit{Energy bands---}As seen in the quasiparticle bands of Fig.~4, the full gap opens in $x$-dominated bands, and the $z$ bonding band is flat and filled. At $T=0$ the indirect gap $G$ is 278 (293) meV in Pmmn (Am2m), both band edges of $x$ character; the $z$ bonding band lies about 0.45 eV below the chemical potential and the antibonding band above 1 eV. At 112 K the $x$ bands overlap while the $z$ bonding band stays filled. Thus $z$ carries over half of the moment but none of the low-energy spectrum.

\begin{figure}[t]
\centering
\includegraphics[width=\columnwidth]{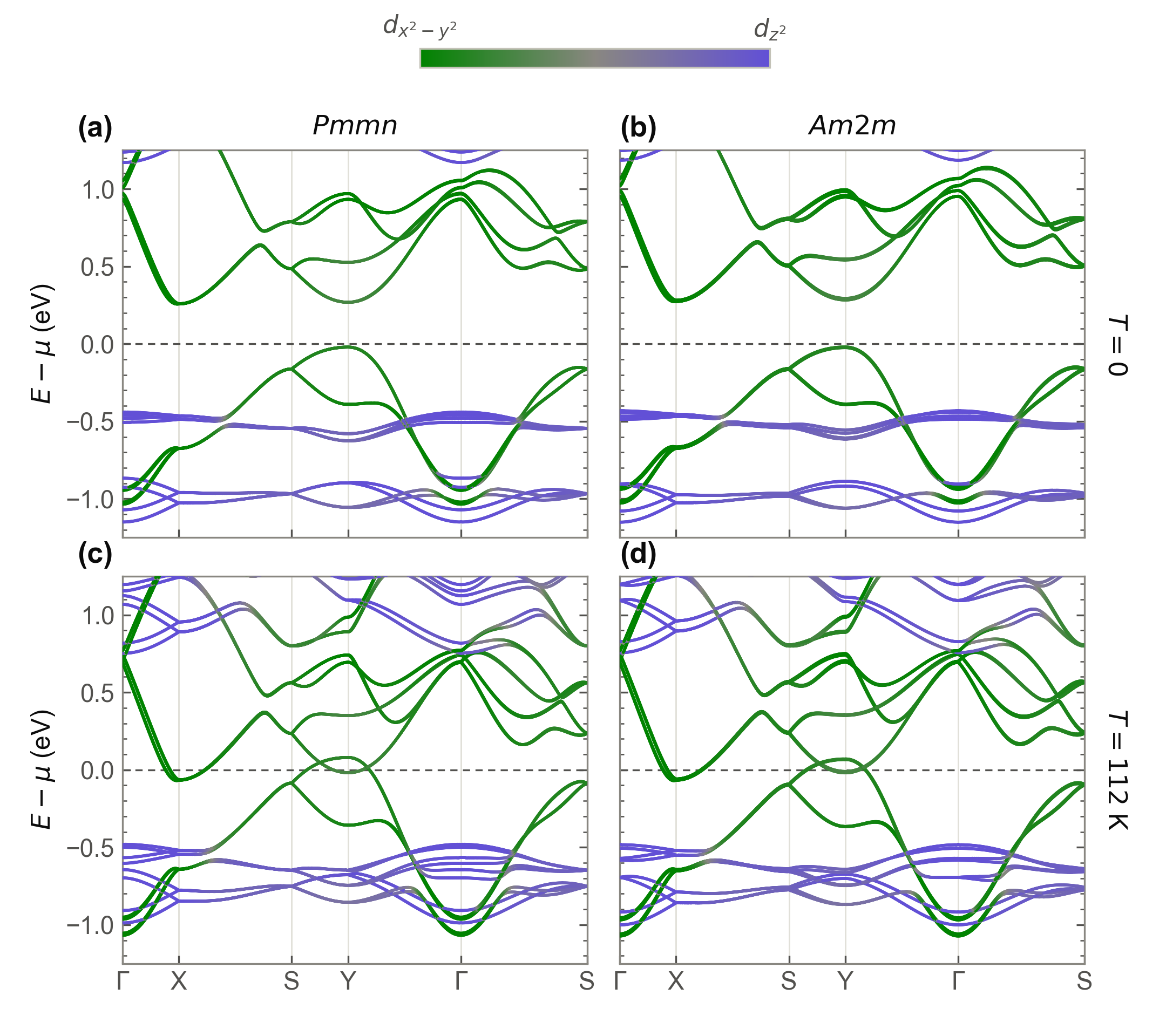}
\caption{Quasiparticle bands of the Pmmn phase (a, c) and the Am2m phase (b, d) at $T=0$ (a, b) and 112 K (c, d), respectively. Color: orbital character, from $x$ (green) to $z$ (purple). Dashed lines: chemical potential.}
\end{figure}

\textit{Energy gap versus temperature---}We find that the calculated energy gap is linear in the charge contrast, $G\simeq\gamma(\Delta n-\Delta n_c)$: straight-line fits give $\gamma=3.8$ eV and $\Delta n_c=0.30$ for the solutions along the warming path in Fig.~5(c), and $\gamma=2.2$ eV and $\Delta n_c=0.25$ for those obtained by varying $V_{\rm rep}$ at $T=0$ in Fig.~5(d). On warming the gap therefore closes continuously while the charge order is still finite, at 106.4 K in Pmmn and 107.9 K in Am2m, as shown in Figs.~5(a) and 5(b). Within 1 K the bands overlap further, and $\Delta n$ and the breathing amplitude $Q^*$ drop together. Above, a metallic stripe state with spin order, charge modulation, and breathing distortion persists up to 143–146 K. The closure of the energy gap and the loss of magnetic order thus occur at two separate temperatures.

\begin{figure}[t]
\centering
\includegraphics[width=\columnwidth]{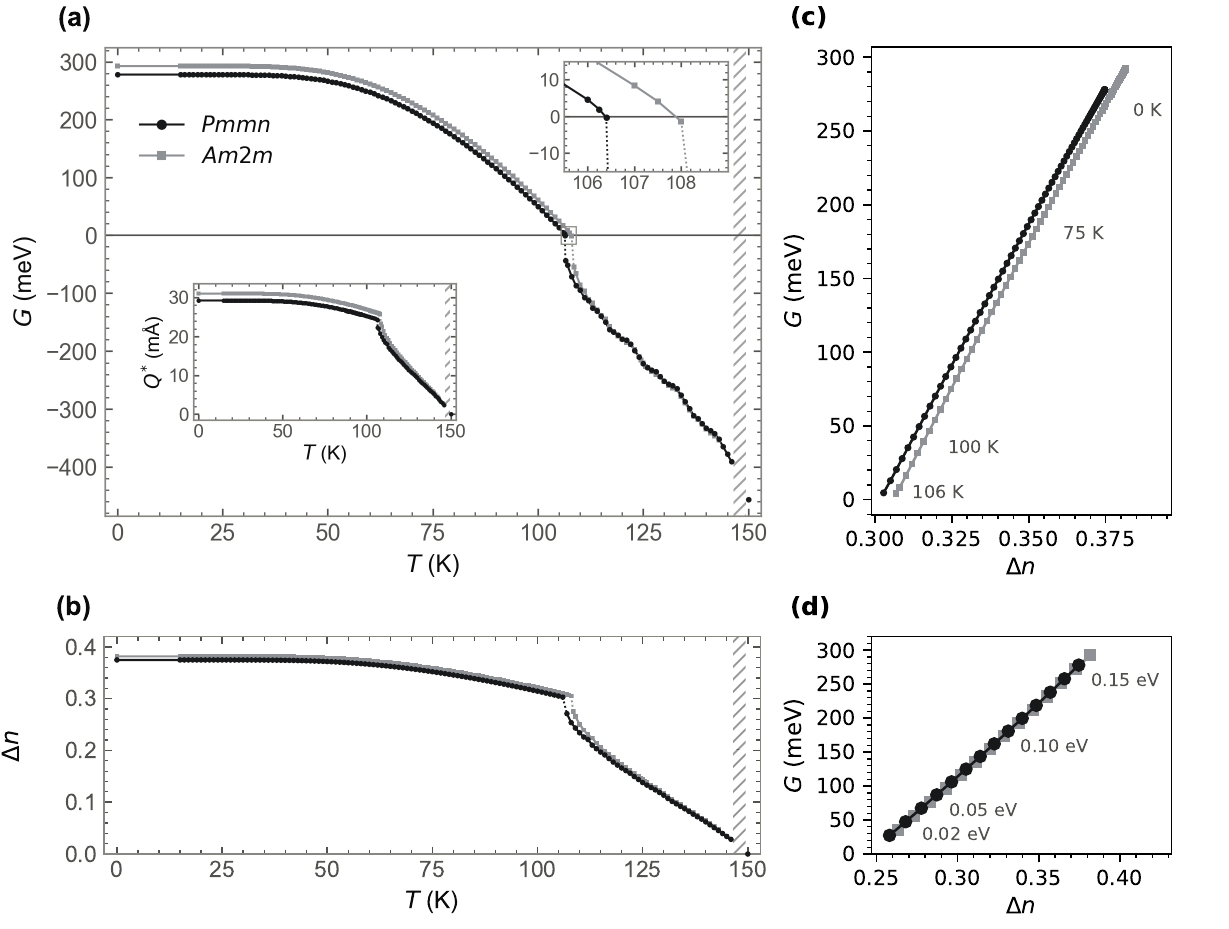}
\caption{(a) Indirect gap $G$ and (b) $\Delta n$ versus temperature; insets: gap closure and breathing amplitude $Q^*$. $G$ versus $\Delta n$ with (c) varying temperature and (d) varying $V_{\rm rep}$ at $T=0$ (labels).}
\end{figure}

\textit{Discussion and conclusion---}Interestingly, we observe that the charge order parameter is proportional to the square of the magnetic moment of the $x$ component of the spin stripe; the charge order and distortion together split the $x$ bands and open the full gap, and the energy gap linearly depends on the charge contrast. On warming, the $x$ moment decreases, the gap closes first, and the stripe with its charge modulation disappears only later. It seems that the charge order is a secondary order of the spin stripe, tied by symmetry to the square of the $x$ moment: it is present as long as the stripe is, and requires neither a Fermi-surface instability nor a gap. The gap is indirect and exists only beyond the threshold $\Delta n_c$ of the charge contrast. As warming lowers the moment continuously, this threshold is crossed before the stripe disappears, and the stripe in between is metallic, with charge order but no gap.

The present results are unified in one physical scenario: the electron correlations induce spin stripe order and charge disproportionation, and the electron–phonon coupling drives the breathing distortions, whose sign in adjacent bilayers selects the polar Am2m or antipolar Pmmn phase; the two $z$ electrons of a dimer form a superposition of bonding-pair and antibonding-pair configurations, with an antibonding-pair weight of 15–22\%, and their interlayer self-energy reaches more than twice the bare interlayer hopping. Obviously, due to Hund's rule coupling, such partially polarized dimers distinctly differ from the $S=0$ singlets.

The polar structure proposed from single-crystal diffraction \cite{ref05} and the Pmnm structure obtained in calculations including $U$ \cite{ZhangB2025PRB}, which shares its space group with Pmmn, thus appear as two minima of the same stripe. The alternation of large and small moments \cite{ref02,ref03}, the separation of the lower-temperature transport and optical anomalies from the onset of magnetic order \cite{ref13,ref15,ref14}, the isotope shift of the lower-temperature anomaly alone \cite{ref16}, and the different orbital characters of the flat band and the gap edges \cite{ref09,ref10} can be read within the same scenario. Here the lower temperature is that of the gap closure, 106–108 K, whereas the charge modulation persists with the stripe up to 143–146 K, so that charge order sets in together with the spin order, as indicated by NQR \cite{LuoJ2025CPL}.

In the cluster RISB method adopted here, all interactions within the dimers are treated exactly, and the motions of the electrons between dimers are treated with the RISB mean-field method, which allows local multiplets and their coherences to vary and retains spatial correlations through its cluster extension \cite{ref31}. However, it is known that the static saddle point of the RISB method overestimates the magnetic moments and the low-temperature gap; meanwhile, the calculation is restricted to collinear order, takes its temperature scale from the experimental $T_{\rm SDW}$, treats $V_{\rm rep}$ as a model parameter, and couples to the lattice through on-site energies only. On the whole, the present method is of high precision, except that the quantum fluctuations between dimers are only partially taken into account.

\bibliography{refs}

\end{document}